\documentclass[pre,twocolumn,amsmath,amssymb,floatfix,superscriptaddress]{revtex4-1}

\usepackage{graphicx}% Include figure files
\usepackage{color}

\begin{document}

\title{Limits of memory coefficient in measuring correlated bursts} % Force line breaks with \\
\author{Hang-Hyun Jo}
\email{hang-hyun.jo@apctp.org}
\affiliation{Asia Pacific Center for Theoretical Physics, Pohang 37673, Republic of Korea}
\affiliation{Department of Physics, Pohang University of Science and Technology, Pohang 37673, Republic of Korea}
\affiliation{Department of Computer Science, Aalto University, Espoo FI-00076, Finland}
\author{Takayuki Hiraoka}
\email{takayuki.hiraoka@apctp.org}
\affiliation{Asia Pacific Center for Theoretical Physics, Pohang 37673, Republic of Korea}

\date{\today}% It is always \today, today,

\begin{abstract}
    Temporal inhomogeneities in event sequences of natural and social phenomena have been characterized in terms of interevent times and correlations between interevent times. The inhomogeneities of interevent times have been extensively studied, while the correlations between interevent times, often called correlated bursts, are far from being fully understood. For measuring the correlated bursts, two relevant approaches were suggested, i.e., memory coefficient and burst size distribution. Here a burst size denotes the number of events in a bursty train detected for a given time window. Empirical analyses have revealed that the larger memory coefficient tends to be associated with the heavier tail of burst size distribution. In particular, empirical findings in human activities appear inconsistent, such that the memory coefficient is close to $0$, while burst size distributions follow a power law. In order to comprehend these observations, by assuming the conditional independence between consecutive interevent times, we derive the analytical form of the memory coefficient as a function of parameters describing interevent time and burst size distributions. Our analytical result can explain the general tendency of the larger memory coefficient being associated with the heavier tail of burst size distribution. We also find that the apparently inconsistent observations in human activities are compatible with each other, indicating that the memory coefficient has limits to measure the correlated bursts.
\end{abstract}

\maketitle

\section{Introduction}\label{sec:intro}

A number of dynamical processes in natural and social phenomena are known to show non-Poissonian or inhomogeneous temporal patterns. Solar flares~\cite{Wheatland1998WaitingTime}, earthquakes~\cite{Corral2004LongTerm, deArcangelis2006Universality, Lippiello2007Dynamical, deArcangelis2016Statistical}, neuronal firings~\cite{Kemuriyama2010Powerlaw}, and human activities~\cite{Barabasi2005Origin, Karsai2018Bursty} are just a few examples. Such temporal inhomogeneities have often been described in terms of $1/f$ noise~\cite{Bak1987Selforganized, Weissman19881f, Ward20071f}. Recently, temporal correlations in event sequences have been studied using the notion of bursts, i.e., rapidly occurring events within short time periods alternating with long inactive periods~\cite{Barabasi2005Origin, Karsai2018Bursty}. It is well-known that bursty interactions between individuals strongly affect the dynamical processes taking place in a network of individuals, such as spreading or diffusion~\cite{Vazquez2007Impact, Karsai2011Small, Miritello2011Dynamical, Rocha2011Simulated, Jo2014Analytically, Delvenne2015Diffusion}. Therefore, it is important to characterize such temporal inhomogeneities or bursts and to understand the underlying mechanisms behind those complex phenomena.

At the simplest level, the bursty dynamics can be characterized by the heavy-tailed interevent time distribution $P(\tau)$, where $\tau$ denotes the time interval between two consecutive events. In many cases, $P(\tau)$ follows a power law with exponent $\alpha$:
\begin{equation}
    \label{eq:Ptau_simple}
    P(\tau)\sim \tau^{-\alpha}.
\end{equation}
The higher-order description of bursts concerns with correlations between interevent times, often called \emph{correlated bursts}~\cite{Karsai2012Universal, Karsai2012Correlated, Jo2015Correlated, Jo2017Modeling}. We find two relevant approaches for studying such correlations between interevent times, i.e., memory coefficient and burst size distribution. For a given sequence of $n$ interevent times, $\{\tau_i\}_{i=1,\cdots,n}$, the memory coefficient is defined as a Pearson correlation coefficient between two consecutive interevent times~\cite{Goh2008Burstiness}:
\begin{equation}
    \label{eq:memory_original}
    M\equiv \frac{ \langle \tau_i\tau_{i+1}\rangle- \langle \tau_i\rangle \langle \tau_{i+1}\rangle }{\sigma_i\sigma_{i+1} },
\end{equation}
where $\langle \tau_i\rangle$ ($\langle\tau_{i+1}\rangle$) and $\sigma_i$ ($\sigma_{i+1}$) denote the average and standard deviation of interevent times except for the last (the first) interevent time, respectively. Positive $M$ implies that large (small) interevent times tend to follow large (small) ones. The opposite tendency is observed for the negative $M$, while $M=0$ indicates no correlations between interevent times. This memory coefficient has been used to analyze event sequences in natural phenomena and human activities as well as to test models for bursty dynamics~\cite{Goh2008Burstiness, Wang2015Temporal, Bottcher2017Temporal, Jo2015Correlated}. For example, it has been found that $M\approx 0.2$ for earthquakes in Japan, while $M$ is close to $0$ or less than $0.1$ for various human activities~\cite{Goh2008Burstiness}. In another work on emergency call records in a Chinese city, individual callers are found to show diverse values of $M$, i.e., a broad distribution of $M$ ranging from $-0.2$ to $0.5$ but peaked at $M=0$~\cite{Wang2015Temporal}. Based on these empirical observations, it appears that most human activities do not show strong correlations between interevent times.

As $M$ measures correlations only between two consecutive interevent times, another approach using the notion of bursty trains was suggested~\cite{Karsai2012Universal}. A bursty train, or burst, is defined as a set of consecutive events for a given time window $\Delta t$, such that interevent times between any two consecutive events in the burst are less than or equal to $\Delta t$, while those between events belonging to different bursts are larger than $\Delta t$. The number of events in the burst is called burst size, denoted by $b$. If the interevent times are fully uncorrelated with each other, the distribution of $b$ follows an exponential function, irrespective of the form of the interevent time distribution. However, the empirical analyses have revealed that the burst size distributions tend to show power-law tails with exponent $\beta$:
\begin{equation}
    \label{eq:burstSizeDistribution}
    Q_{\Delta t}(b)\sim b^{-\beta} 
\end{equation}
for a wide range of $\Delta t$, e.g., in earthquakes, neuronal activities, and human communication patterns~\cite{Karsai2012Universal, Karsai2012Correlated, Wang2015Temporal}. For example, the empirical value of $\beta$ varies from $2.5$ for earthquakes in Japan to $3.9$--$4.2$ for mobile phone communication patterns~\cite{Karsai2012Universal, Karsai2012Correlated}, while it is found that $\beta\approx 2.21$ in the emergency call dataset~\cite{Wang2015Temporal}. Such power-law burst size distributions for a wide range of $\Delta t$ may indicate that there exists a hierarchical burst structure~\footnote{We also note that the exponential burst size distributions have been reported for mobile phone calls of individual users in another work~\cite{Jiang2016Twostate}. These inconsistent results for burst size distributions raise a debatable issue about the existence of the hierarchical burst structure in human communication patterns. This is however beyond the scope in this paper.}, which however seems to be inconsistent with the observation of $M\approx 0$ in human activities because $M\approx 0$ implies negligible correlations between interevent times. We also observe a general tendency that the larger value of $M$ is associated with the smaller value of $\beta$, which can be understood by the intuition that the smaller $\beta$ implies the stronger correlations between interevent times, possibly leading to the larger $M$. However, little is known about the relation between memory coefficient and burst size distribution, requiring us to rigorously investigate their relation.

In order to systematically study the relation between memory coefficient and burst size distribution, in this paper we derive the analytical form of the memory coefficient as a function of parameters describing interevent time and burst size distributions by assuming the conditional independence between consecutive interevent times. Our analytical result turns out to explain the general tendency that the larger $M$ is associated with the smaller $\beta$. We also find that the apparently inconsistent observations in human activities, i.e., $M\approx 0$ but $Q_{\Delta t}(b)\sim b^{-\beta}$ with $\beta\approx 4$, can be compatible with each other. This finding raises an important question about the effectiveness or limits of the memory coefficient in measuring correlated bursts.

Our paper is organized as follows: In Sec.~\ref{sec:single}, we derive the analytical form of the memory coefficient in the case when a single timescale is used for identifying bursty trains, which is also numerically demonstrated. Then we extend the single timescale analysis to the more realistic case with multiple timescales in Sec.~\ref{sec:multi}. Finally, we conclude our work in Sec.~\ref{sec:concl}.

\section{Single timescale analysis}\label{sec:single}

We analytically study the relation between memory coefficient in Eq.~(\ref{eq:memory_original}) and burst size distribution for a given interevent time distribution, by deriving an analytical form of the memory coefficient as a function of parameters of interevent time and burst size distributions. Here we consider bursty trains detected using one time window or timescale, namely, a single timescale analysis.

\subsection{Analytical derivation of $M$}\label{subsec:single-anal}

\begin{figure}[!t]
    \includegraphics[width=\columnwidth]{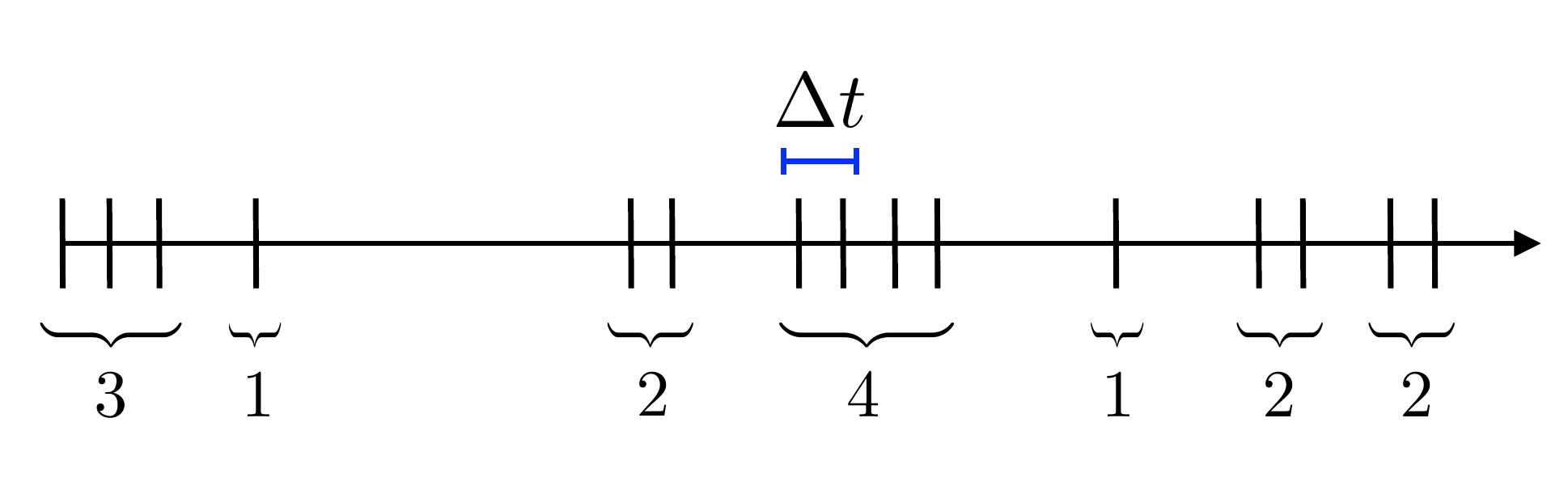}
    \caption{An example of bursty trains detected using a time window $\Delta t$ for an event sequence in time. Each vertical line denotes an event, and the numbers indicate the sizes of bursty trains.}
    \label{fig:example}
\end{figure}

Let us assume that an event sequence with $n+1$ events is characterized by $n$ interevent times, denoted by $T\equiv\{\tau_1,\cdots, \tau_n\}$, and that for a given $\Delta t$ one can detect $m$ bursty trains whose sizes are denoted by $B\equiv\{b_1,\cdots, b_m\}$, see Fig.~\ref{fig:example} for an example. The sum of burst sizes must be the number of events, i.e., $\sum_{j=1}^m b_j=n+1$. With $\langle b\rangle$ denoting the average burst size, we can write
\begin{equation}
    \label{eq:mn_event}
    m\langle b\rangle = n+1 \simeq n,
\end{equation}
where the approximation has been made in the asymptotic limit with $n\gg 1$. The number of bursty trains is related to the number of interevent times larger than $\Delta t$, i.e., 
\begin{equation}
    \label{eq:mn_condition}
    m=|\{\tau_i |\tau_i>\Delta t\}|+1. 
\end{equation}
In the asymptotic limit with $n,m\gg 1$, we get 
\begin{equation}
    \label{eq:mn_burst}
    m\simeq n\Pr(\tau>\Delta t).
\end{equation}
By combining Eqs.~(\ref{eq:mn_event}) and (\ref{eq:mn_burst}), we obtain a general relation as
\begin{equation}
    \label{eq:mn_event_burst}
    \langle b\rangle \Pr(\tau>\Delta t) \simeq 1,
\end{equation}
which holds for arbitrary functional forms of interevent time and burst size distributions~\cite{Jo2017Modeling}. These distributions will be denoted by $P(\tau)$ and $Q_{\Delta t}(b)$, respectively. 

We now derive the memory coefficient: Using a given $\Delta t$, we divide $T$ into two subsets as
\begin{eqnarray}
    \label{eq:T_subsets}
    T_0 &\equiv & \{\tau_i| \tau_i\leq \Delta t\},\\
    T_1 &\equiv & \{\tau_i| \tau_i> \Delta t\}.
\end{eqnarray}
The set of all pairs of two consecutive interevent times, $\{(\tau_i,\tau_{i+1})\}$, can be divided into four subsets as follows:
\begin{equation}
    \label{eq:subsets_pair}
    T_{\mu\nu} \equiv \{(\tau_i,\tau_{i+1})|\tau_i\in T_\mu, \tau_{i+1}\in T_\nu\},
\end{equation}
where $\mu,\nu \in \{0,1\}$. By denoting the fraction of interevent time pairs in each $T_{\mu\nu}$ by $t_{\mu\nu}\equiv \langle |T_{\mu\nu}|\rangle / (n-1)$, the term $\langle\tau_i\tau_{i+1}\rangle$ in Eq.~(\ref{eq:memory_original}) can be written as
\begin{equation}
    \langle \tau_i\tau_{i+1}\rangle = \sum_{\mu,\nu\in \{0,1\}} t_{\mu\nu} \tau^{(\mu)} \tau^{(\nu)},
\end{equation}
where
\begin{equation}
    \tau^{(0)} \equiv \frac {\int_0^{\Delta t}\tau P(\tau)d\tau} {\int_0^{\Delta t}P(\tau)d\tau},\ 
    \tau^{(1)} \equiv \frac {\int_{\Delta t}^\infty \tau P(\tau)d\tau} {\int_{\Delta t}^\infty P(\tau)d\tau}.
    \label{eq:tau01_mean}
\end{equation}
Here we have assumed that the information on the correlation between $\tau_i$ and $\tau_{i+1}$ is carried only by $t_{\mu\nu}$, while such consecutive interevent times are independent of each other under the condition that $\tau_i\in T_\mu$ and $\tau_{i+1}\in T_\nu$. This assumption of conditional independence is based on the fact that the correlation between $\tau_i$ and $\tau_{i+1}$ with $\tau_i\in T_\mu$ and $\tau_{i+1}\in T_\nu$ is no longer relevant to the burst size statistics, because the bursty trains are determined depending only on whether each interevent time is larger than $\Delta t$ or not. Then $M$ in Eq.~(\ref{eq:memory_original}) reads in the asymptotic limit with $n\gg 1$
\begin{equation}
    \label{eq:memory_approx}
    M\simeq \frac{ \sum_{\mu,\nu\in \{0,1\}} t_{\mu\nu} \tau^{(\mu)} \tau^{(\nu)} - \langle \tau\rangle^2}{\sigma^2}.
\end{equation}
Here we have approximated as $\langle \tau_i\rangle \simeq \langle \tau_{i+1}\rangle \simeq \langle \tau\rangle$ and $\sigma_i \simeq \sigma_{i+1} \simeq \sigma$, with $\langle\tau\rangle$ and $\sigma$ denoting the average and standard deviation of interevent times, respectively. Note that $\tau^{(0)}$ and $\tau^{(1)}$ are related as follows:
\begin{equation}
    \label{eq:tau01_relation}
    \left(1-\frac{1}{\langle b\rangle}\right) \tau^{(0)} +\frac{1}{\langle b\rangle} \tau^{(1)} \simeq \langle \tau\rangle.
\end{equation}

For deriving $M$ in Eq.~(\ref{eq:memory_approx}), one needs to calculate $t_{\mu\nu}$. Since each pair of interevent times in $T_{11}$ implies a burst of size $1$, the average size of $T_{11}$ is $mQ_{\Delta t}(1)$. Thus, the average fraction of interevent time pairs in $T_{11}$ becomes
\begin{equation}
    \label{eq:t11_single}
    t_{11}\equiv \frac{\langle |T_{11}|\rangle}{n-1} \simeq \frac{Q_{\Delta t}(1)}{\langle b\rangle},
\end{equation}
where Eq.~(\ref{eq:mn_event}) has been used. The pair of interevent times in $T_{10}$ ($T_{01}$) is found whenever a burst of size larger than $1$ begins (ends). Hence, the average fraction of $T_{10}$, equivalent to that of $T_{01}$, must be 
\begin{equation}
    \label{eq:t10_single}
    t_{10} \equiv \frac{\langle |T_{10}|\rangle}{n-1} \simeq \frac{1}{\langle b\rangle}\sum_{b=2}^\infty Q_{\Delta t}(b) = \frac{1-Q_{\Delta t}(1)}{\langle b\rangle},
\end{equation}
which is the same as $t_{01} \equiv \langle |T_{01}|\rangle/(n-1)$. Finally, for each burst of size larger than $2$, we find $b-2$ pairs of interevent times belonging to $T_{00}$, indicating that the average fraction of $T_{00}$ is
\begin{equation}
    \label{eq:t00_single}
    t_{00} \equiv \frac{\langle |T_{00}|\rangle}{n-1} \simeq \frac{1}{\langle b\rangle}\sum_{b=3}^\infty (b-2)Q_{\Delta t}(b) = \frac{\langle b\rangle -2 +Q_{\Delta t}(1)}{\langle b\rangle}.
\end{equation}
Note that $t_{00} + t_{01} + t_{10} + t_{11}\simeq 1$. Then by using Eqs.~(\ref{eq:tau01_mean}) and~(\ref{eq:tau01_relation}) one obtains
\begin{equation}
    \sum_{\mu,\nu\in \{0,1\}} t_{\mu\nu}\tau^{(\mu)} \tau^{(\nu)} = [\langle b\rangle Q_{\Delta t}(1)-1] ( \langle \tau\rangle - \tau^{(0)})^2 +\langle \tau\rangle^2, 
\end{equation}
leading to
\begin{equation}
    \label{eq:memory_result}
    M \simeq \frac{[\langle b\rangle Q_{\Delta t}(1)-1] ( \langle \tau\rangle - \tau^{(0)})^2}{\sigma^2}.
\end{equation}
Note that this analytical result has been obtained for arbitrary functional forms of interevent time and burst size distributions. It turns out that the value of $Q_{\Delta t}(1)$, i.e., the fraction of bursts consisting of standalone events, also plays an important role in determining the value of $M$.

We investigate the dependence of $M$ on $Q_{\Delta t}(b)$, while keeping the same $P(\tau)$. As for the burst size distribution, we consider a power-law distribution as follows:
\begin{equation}
    \label{eq:Qb_powerlaw}
    Q_{\Delta t}(b) = \zeta(\beta)^{-1} b^{-\beta}\ \textrm{for}\ b=1,2,\cdots,
\end{equation}
where $\zeta(\cdot)$ denotes the Riemann zeta function. We assume that $\beta>2$ for the existence of $\langle b\rangle$, i.e., $\langle b\rangle = \zeta(\beta-1)/\zeta(\beta)$. As for the interevent time distribution, a power-law distribution with an exponential cutoff is considered:
\begin{equation}
    \label{eq:Ptau_powerlaw}
    P(\tau)= \left\{\begin{tabular}{ll}
            $C\tau^{-\alpha}e^{-\tau/\tau_c}$ & for $\tau\geq \tau_{\rm min}$,\\
            $0$ & otherwise,
        \end{tabular}\right.
\end{equation}
where $\tau_{\rm min}$ and $\tau_c$ denote the lower bound and the exponential cutoff of $\tau$, respectively, and $C\equiv\tau_c^{\alpha-1}/\Gamma(1-\alpha,\tau_{\rm min}/\tau_c)$ is the normalization constant. Here $\Gamma(\cdot,\cdot)$ denotes the upper incomplete Gamma function.

\begin{figure}[!t]
    \includegraphics[width=.75\columnwidth]{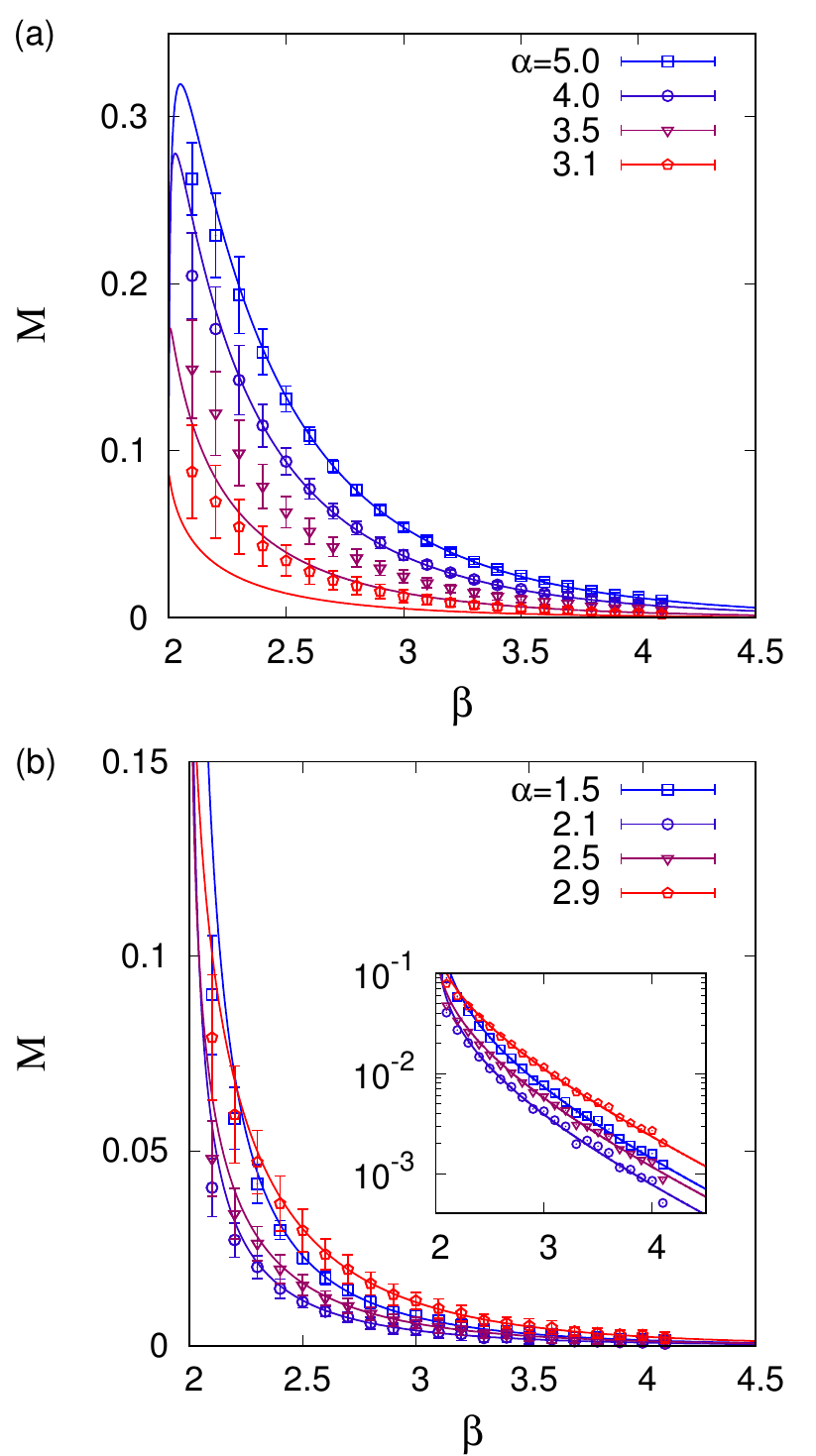}
    \caption{The analytical solution of $M$ in Eq.~(\ref{eq:memory_result}) as a function of $\beta$ in Eq.~(\ref{eq:Qb_powerlaw}) for several values of $\alpha$ in Eq.~(\ref{eq:Ptau_powerlaw}) (solid lines), compared with corresponding numerical results (symbols with error bars). In panel (a) we use the pure power-law distribution of $P(\tau)$ in Eq.~(\ref{eq:Ptau_powerlaw}), with infinite exponential cutoff, i.e., $\tau_c\to\infty$, while the general form of $P(\tau)$ with $\tau_c=10^3\tau_{\rm min}$ is used in panel (b). The inset shows the same result as in panel (b), but in a semi-log scale. Each point and its standard deviation are obtained from $50$ event sequences of size $n=5\times 10^5$.}
    \label{fig:single}
\end{figure}

We first consider the pure power-law case of $P(\tau)$ with $\tau_c\to\infty$, where $\alpha>3$ is assumed for the existence of $\sigma$. From Eq.~(\ref{eq:mn_event_burst}), we obtain the relation between parameters as
\begin{equation}
    \label{eq:relation_param}
    \langle b\rangle = \frac{\zeta(\beta-1)}{\zeta(\beta)} \simeq \left(\frac{\Delta t}{\tau_{\rm min}}\right)^{\alpha-1}.
\end{equation}
In order to study the dependence of $M$ on $\beta$, we fix the values of $\alpha$ and $\tau_{\rm min}$ for keeping the same $P(\tau)$, implying that $\langle\tau\rangle$ and $\sigma$ in Eq.~(\ref{eq:memory_result}) remain the same. Then the variation of $\beta$ affects only $\Delta t$ by means of Eq.~(\ref{eq:relation_param}), consequently $\tau^{(0)}$ in Eq.~(\ref{eq:memory_result}) as
\begin{equation}
    \tau^{(0)} = \frac {(\alpha-1)[1-(\Delta t/\tau_{\rm min})^{2-\alpha}]} {(\alpha-2)[1-(\Delta t/\tau_{\rm min})^{1-\alpha}]} \tau_{\rm min}.
\end{equation}
In our setting, $\Delta t$ is not a control parameter but it is automatically determined by other parameters, i.e., $\alpha$, $\tau_{\rm min}$, and $\beta$~\footnote{Although the power-law tails of burst size distributions have been shown to be robust with respect to the variation of $\Delta t$ in several empirical analyses, the value of $\Delta t$ might be related to a specific timescale in some phenomena. In such cases, more realistic approach should be taken so that both $\beta$ and $\Delta t$ are control parameters, hence one can study the effect of $\beta$ on $M$ without bothering with the choice of $\Delta t$.}. The stronger correlations between interevent times can be characterized by the smaller value of $\beta$, i.e., the larger $\langle b\rangle$ and the smaller $Q_{\Delta t}(1)$. It also leads to the larger $\Delta t$ by means of Eq.~(\ref{eq:relation_param}), hence the larger $\tau^{(0)}$ and the smaller $\langle \tau\rangle - \tau^{(0)}$. As it is not straightforward to see whether $M$ is increasing or decreasing according to $\beta$, we plot the analytical result of $M$ in Eq.~(\ref{eq:memory_result}) for various values of parameters, as depicted by solid lines in Fig.~\ref{fig:single}(a). We find that $M$ is an overall decreasing function of $\beta$ as expected, implying that the stronger correlations between interevent times, i.e., the smaller $\beta$, lead to the larger value of $M$. In the limiting case with $\beta\to \infty$, $\Delta t$ approaches $\tau_{\rm min}$, implying that the interevent times are rarely correlated with each other, and hence $M\to 0$. In addition, for the sufficiently large $\alpha$, $M$ turns out to increase according to $\beta$ in the vicinity of $\beta=2$: The excessively strong correlations between interevent times can even reduce the value of the memory coefficient. Finally, we also find that the smaller $\alpha$ leads to the smaller $M$ for a fixed $\beta$, which can be understood by the fact that the large $\Delta t$ due to the small $\alpha$ enhances the mixing of interevent times with various timescales within $T_0$.

Next, we consider the general form of $P(\tau)$ in Eq.~(\ref{eq:Ptau_powerlaw}) with finite $\tau_c$, allowing us to study a more realistic, wider range of $\alpha$ observed in the empirical analyses, e.g., in Ref.~\cite{Karsai2018Bursty}. Once $\Delta t$ is determined from the relation 
\begin{equation}
    \langle b\rangle = \frac{\zeta(\beta-1)}{\zeta(\beta)} \simeq 
        \frac{\Gamma(1-\alpha, \tau_{\rm min}/\tau_c)}{\Gamma(1-\alpha, \Delta t/\tau_c)}
\end{equation}
for given $\alpha$, $\tau_{\rm min}$, $\tau_c$, and $\beta$, the calculation of $M$ is straightforward by using
\begin{eqnarray}
    \langle \tau\rangle &=& \frac{\Gamma(2-\alpha, \tau_{\rm min}/\tau_c)}{\Gamma(1-\alpha, \tau_{\rm min}/\tau_c)} \tau_c,\\
    \tau^{(0)} &=& \frac{\Gamma(2-\alpha, \tau_{\rm min}/\tau_c) - \Gamma(2-\alpha, \Delta t/\tau_c)}{\Gamma(1-\alpha, \tau_{\rm min}/\tau_c) - \Gamma(1-\alpha, \Delta t/\tau_c)} \tau_c.
\end{eqnarray}
The analytical result of $M$ in Eq.~(\ref{eq:memory_result}) is plotted for various values of parameters, as depicted by solid lines in Fig.~\ref{fig:single}(b). Here we have used $\tau_c=10^3\tau_{\rm min}$. It is found that $M$ is a decreasing function of $\beta$ as expected, implying that the stronger correlations between interevent times lead to the larger value of $M$. For a fixed value of $\beta$, $M$ shows non-monotonic behaviors according to $\alpha$, which could be related to the non-monotonic behaviors of the decaying exponent of autocorrelation function as a function of $\alpha$, as reported in Ref.~\cite{Jo2017Modeling}. More importantly, the value of $M$ turns out to be much smaller than $0.1$ for the wide range of $\beta$. In particular, we find $M$ close to $0$ for $\beta\approx 4$, regardless of the value of $\alpha$. Hence, $M\approx 0$ does not necessarily mean no correlations between interevent times. This result can help us resolve the issue regarding the apparently conflicting observations in the mobile phone datasets, i.e., $M\approx 0$ but $Q_{\Delta t}(b)\sim b^{-\beta}$ with $\beta\approx 4$~\cite{Goh2008Burstiness, Karsai2012Universal}. These observations can be compatible with each other.

\subsection{Numerical demonstration}\label{subsec:single-numeric}

In order to numerically demonstrate the effect of the burst size distribution on the memory coefficient, we adopt the method suggested for implementing correlated bursts~\cite{Jo2017Modeling}: We prepare $n$ uncorrelated interevent times that are independently drawn from $P(\tau)$ in Eq.~(\ref{eq:Ptau_powerlaw}), which is denoted by $T=\{\tau_1,\cdots,\tau_n\}$. Then burst sizes are independently drawn from $Q_{\Delta t}(b)$ in Eq.~(\ref{eq:Qb_powerlaw})~\footnote{For the generation of the power-law distribution of discrete values, we referred to the method in Ref.~\cite{Clauset2009Powerlaw}.} one by one until the sum of burst sizes exceeds $n+1$. Once the sum of burst sizes exceeds $n+1$, the last burst size is reduced by the excessive amount so that the sum of burst sizes becomes exactly the same as $n+1$. Then the resultant number of burst sizes is denoted by $m$, hence $B=\{b_1,\cdots, b_m\}$ with $\sum_{j=1}^m b_j = n+1$. Note that the value of $\Delta t$ is automatically determined by Eq.~(\ref{eq:mn_condition}) for given $T$ and $m$. Using this $\Delta t$, we divide $T$ into two subsets as $T_0 =\{\tau_i| \tau_i\leq \Delta t\}$ and $T_1 =\{\tau_i| \tau_i> \Delta t\}$. Next, in order to implement the correlations between interevent times, one can permute or reconstruct the interevent times in $T$ according to $B$. Let us prepare an empty sequence for correlated interevent times, $T'$. We first randomly draw a burst size, say $b$, from $B$ without replacement. If $b>1$, we randomly draw $b-1$ interevent times from $T_0$ without replacement and one interevent time from $T_1$ without replacement. Otherwise, if $b=1$, we randomly draw one interevent time from $T_1$ without replacement. These $b$ interevent times are sequentially added to $T'$. Then another burst size is randomly drawn from $B$ and the same process is repeated until all burst sizes in $B$ as well as all interevent times in $T_0$ and $T_1$ are used up, i.e., until $|T'|=n$. Once the sequence of interevent times, $T'$, is obtained, we immediately calculate the value of $M$ in Eq.~(\ref{eq:memory_original}).

The numerical results of $M$ as a function of $\beta$ for various values of $\alpha$ are shown in Fig.~\ref{fig:single}, where each point is obtained from $50$ event sequences of size $n=5\times 10^5$. For both cases with and without exponential cutoffs for $P(\tau)$ in Eq.~(\ref{eq:Ptau_powerlaw}), we find that the numerical results are comparable to the analytical values of $M$. However, systematic deviations are observed in the pure power-law case, especially for small values of $\alpha$ and $\beta$, where the natural cutoffs of power-law distributions, i.e., $P(\tau)$ and/or $Q_{\Delta t}(b)$, become effective due to the finite sizes of $n$ and/or $m$. In addition, the increasing behavior of $M$ according to $\beta$ for the region of large $\alpha$ and small $\beta$ turns out to be rarely visible from our numerical simulations. In the case with power-law with exponential cutoff, we also find relatively larger deviations of $M$ for the range of $\beta\approx 2$ probably due to the similar finite size effects as mentioned.

\section{Multiple timescale analysis}\label{sec:multi}

In order to study more realistic cases, i.e., power-law burst size distributions for a wide range of time windows~\cite{Karsai2012Universal, Karsai2012Correlated, Wang2015Temporal}, we extend the single timescale analysis in the previous Section to a multiple timescale case, where more than one time window, $\Delta t$, are used for detecting bursty trains at multiple timescales.

\subsection{Analytical derivation of $M$}\label{subsec:multi-anal}

As the simplest case, we apply two timescales or time windows, i.e., $\Delta t_0$ and $\Delta t_1$ with $\Delta t_0<\Delta t_1$, to the set of $n$ interevent times, denoted by $T\equiv\{\tau_1,\cdots, \tau_n\}$. Then for a given $\Delta t_l$ ($l=0,1$), one can detect $m_l$ bursty trains whose sizes are denoted by $B_l\equiv\{b^{(l)}_1,\cdots, b^{(l)}_{m_l}\}$. The sum of burst sizes must be the number of events, i.e., $\sum_{j=1}^{m_l} b^{(l)}_j=n+1$ for each $l$. In the asymptotic limit with $n\gg 1$, we can write
\begin{equation}
    \label{eq:mn_event_multi}
    m_l\langle b_l\rangle \simeq n,
\end{equation}
where $\langle b_l\rangle \equiv \langle b^{(l)}\rangle$ denotes the average burst size when using $\Delta t_l$. The number of bursty trains is related to the number of interevent times larger than $\Delta t_l$, i.e., 
\begin{equation}
    \label{eq:mn_burst_multi}
    m_l\simeq n\Pr(\tau>\Delta t_l).
\end{equation}
By combining Eqs.~(\ref{eq:mn_event_multi}) and (\ref{eq:mn_burst_multi}), we obtain a general relation for each $l$ as
\begin{equation}
    \label{eq:mn_event_burst_multi}
    \langle b_l\rangle \Pr(\tau>\Delta t_l) \simeq 1,
\end{equation}
which holds again for arbitrary functional forms of interevent time and burst size distributions~\cite{Jo2017Modeling}. The burst size distributions will be denoted by $Q_{\Delta t_l}(b^{(l)})$, or simply $Q_l(b)$.

The memory coefficient can be derived for $T$, $B_0$, and $B_1$. Using given $\Delta t_0$ and $\Delta t_1$, we divide $T$ into three subsets as
\begin{eqnarray}
    \label{eq:T_subsets_multi}
    T_0 &\equiv & \{\tau_i| \tau_i\leq \Delta t_0\},\\
    T_1 &\equiv & \{\tau_i| \Delta t_0< \tau_i\leq \Delta t_1\},\\
    T_2 &\equiv & \{\tau_i| \tau_i> \Delta t_1\}.
\end{eqnarray}
The set of all pairs of two consecutive interevent times, $\{(\tau_i,\tau_{i+1})\}$, can be divided into nine subsets as follows:
\begin{equation}
    \label{eq:subsets_pair_multi}
    T_{\mu\nu} \equiv \{(\tau_i,\tau_{i+1})|\tau_i\in T_\mu, \tau_{i+1}\in T_\nu\},
\end{equation}
where $\mu,\nu \in \{0,1,2\}$. We then define the fraction of interevent time pairs in each subset as $t_{\mu\nu}\equiv \langle |T_{\mu\nu}|\rangle /(n-1)$, which must be normalized as
\begin{equation}
    \label{eq:t_norm}
    \sum_{\mu,\nu\in \{0,1,2\}} t_{\mu\nu}=1.
\end{equation}
Then one can approximate $M$ in Eq.~(\ref{eq:memory_original}) by  
\begin{equation}
    \label{eq:memory_approx_multi}
    M\simeq \frac{ \sum_{\mu,\nu\in \{0,1,2\}} t_{\mu\nu} \tau^{(\mu)} \tau^{(\nu)} - \langle \tau\rangle^2}{\sigma^2}
\end{equation}
with the average interevent times defined as
\begin{equation}
    \label{eq:tau01_mean_multi}
    \tau^{(0)} \equiv \frac {\int_0^{\Delta t_0}\tau P(\tau)d\tau} {\int_0^{\Delta t_0}P(\tau)d\tau},\
    \tau^{(1)} \equiv \frac {\int_{\Delta t_0}^{\Delta t_1}\tau P(\tau)d\tau} {\int_{\Delta t_0}^{\Delta t_1}P(\tau)d\tau},
\end{equation}
\begin{equation}
    \label{eq:tau2_mean_multi}
    \tau^{(2)} \equiv \frac {\int_{\Delta t_1}^\infty \tau P(\tau)d\tau} {\int_{\Delta t_1}^\infty P(\tau)d\tau},
\end{equation}
satisfying that
\begin{equation}
    \left(1-\frac{1}{\langle b_0\rangle}\right) \tau^{(0)} +\left(\frac{1}{\langle b_0\rangle}- \frac{1}{\langle b_1\rangle} \right) \tau^{(1)} +\frac{1}{\langle b_1\rangle} \tau^{(2)} \simeq \langle \tau\rangle.
\end{equation}

To complete the derivation of $M$, we calculate $t_{\mu\nu}$ for $\mu,\nu\in\{0,1,2\}$ in terms of $Q_l(b)$ for $l=0,1$, in a similar way as done in Sec.~\ref{sec:single}. However, if $\Delta t_0$ is used for detecting bursty trains, interevent times in $T_1$ are not distinguishable from those in $T_2$ as both are considered larger than $\Delta t_0$. Due to this ambiguity, we first define the following quantities, similarly to Eqs.~(\ref{eq:t11_single}--\ref{eq:t00_single}):
\begin{eqnarray}
    c_{00} &\equiv& \frac{\langle b_0\rangle -2 +Q_0(1)}{\langle b_0\rangle},\\
    c_{0*} &=& c_{*0} \equiv \frac{1-Q_0(1)}{\langle b_0\rangle},\\
    c_{**} &\equiv& \frac{Q_0(1)}{\langle b_0\rangle},
\end{eqnarray}
where the subscripts $0$ and $*$ denote $\tau\in T_0$ and $\tau \in T_1\cup T_2$, respectively. These quantities satisfy $c_{00} +c_{0*} +c_{*0} +c_{**}=1$, and they are respectively related with $t_{\mu\nu}$ as
\begin{eqnarray}
    \label{eq:c00}
    c_{00} &\simeq& t_{00},\\
    c_{0*} &\simeq& t_{01} + t_{02},\\
    c_{*0} &\simeq& t_{10} + t_{20},\\
    c_{**} &\simeq& t_{11} + t_{12} + t_{21} + t_{22}.
    \label{eq:c11}
\end{eqnarray}
Similarly, if $\Delta t_1$ is used for detecting bursty trains, interevent times in $T_0$ and $T_1$ are indistinguishable, requiring us to define the following quantities and to relate them to $t_{\mu\nu}$:
\begin{eqnarray}
    \label{eq:d1}
    d_{**} &\equiv& \frac{\langle b_1\rangle -2 +Q_1(1)}{\langle b_1\rangle},\\
    \label{eq:d2}
    d_{*2} &=& d_{2*} \equiv \frac{1-Q_1(1)}{\langle b_1\rangle},\\
    \label{eq:d3}
    d_{22} &\equiv& \frac{Q_1(1)}{\langle b_1\rangle},
\end{eqnarray}
and
\begin{eqnarray}
    \label{eq:d11}
    d_{**} &\simeq& t_{00} + t_{01} + t_{10} + t_{11},\\
    d_{*2} &\simeq& t_{02} + t_{12},\\
    d_{2*} &\simeq& t_{20} + t_{21},\\
    d_{22} &\simeq& t_{22},
    \label{eq:d22}
\end{eqnarray}
where the subscripts $*$ and $2$ denote $\tau\in T_0\cup T_1$ and $\tau \in T_2$, respectively. Note that $d_{**} +d_{*2} +d_{2*} +d_{22}=1$. Although for $\mu\neq\nu$, $t_{\mu\nu}\neq t_{\nu\mu}$ in general, we assume that $t_{\mu\nu}=t_{\nu\mu}$ in the asymptotic limit with $n\gg 1$, leaving $6$ unknowns, i.e., $t_{00}$, $t_{01}$, $t_{02}$, $t_{11}$, $t_{12}$, and $t_{22}$. However, we have only $5$ distinct relations for $t_{\mu\nu}$ as other relations can be simply derived from the normalization condition in Eq.~(\ref{eq:t_norm}):
\begin{eqnarray}
    \label{eq:t00_multi}
    t_{00} &\simeq& c_{00},\\
    t_{01} &\simeq& \frac{d_{**}-c_{00}-t_{11}}{2},\\
    t_{02} &\simeq& c_{0*} - \frac{d_{**}-c_{00}-t_{11}}{2},\\
    t_{12} &\simeq& \frac{c_{**}-d_{22}-t_{11}}{2},\\
    t_{22} &\simeq& d_{22}.
    \label{eq:t22_multi}
\end{eqnarray}
As a result, $t_{\mu\nu}$ cannot be fully determined by $Q_l(b)$ for $l=0,1$.

\begin{figure*}[!t]
    \includegraphics[width=\textwidth]{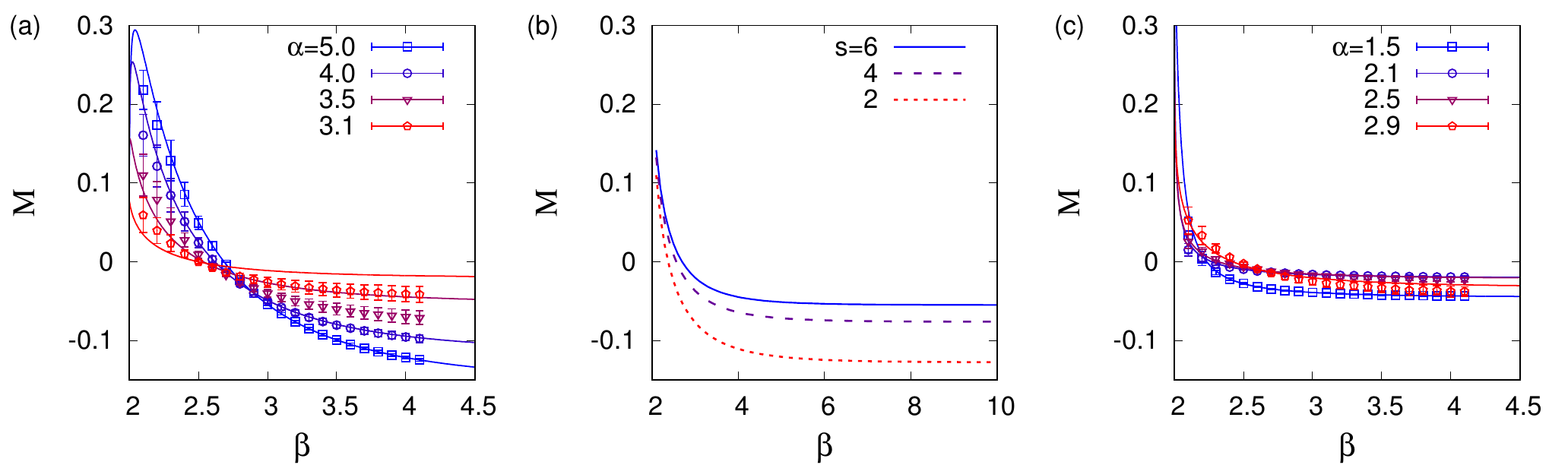}
    \caption{(a) The analytical solution of $M$ in Eq.~(\ref{eq:memory_approx_multi}) as a function of $\beta$ in Eq.~(\ref{eq:Q0b}) for several values of $\alpha$ in Eq.~(\ref{eq:Ptau_powerlaw}) and $s=4$ in Eq.~(\ref{eq:Q1b}) (solid lines), compared with corresponding numerical results (symbols with error bars). We use the pure power-law distribution of $P(\tau)$ in Eq.~(\ref{eq:Ptau_powerlaw}), with infinite exponential cutoff, i.e., $\tau_c\to\infty$. (b) We show the dependence of the analytical solution of $M$ on the parameter $s$ for a fixed value of $\alpha=3.5$. (c) The same as (a) but for the general form of $P(\tau)$ with $\tau_c=10^3\tau_{\rm min}$. In panels (a) and (c), each point and its standard deviation are obtained from $50$ event sequences of size $n=5\times 10^5$.}
    \label{fig:multi}
\end{figure*}

In order to determine $t_{\mu\nu}$, we need to exploit more detailed information on the relation between $Q_0(b)$ and $Q_1(b)$. These two burst size distributions are not independent of each other: For any event sequence, one bursty train detected using $\Delta t_1$ typically consists of more than one bursty train detected using $\Delta t_0$. In other words, one burst size, $b^{(1)}$, at a larger timescale is given as the sum of more than one burst size, $b^{(0)}$, at a smaller timescale. Precisely, for a given $B_0$, burst sizes in $B_0$ are merged to derive burst sizes in $B_1$. Different merging methods can result in different $B_1$, i.e., $Q_1(b)$, from the same $B_0$, i.e., $Q_0(b)$. Here we require both $Q_0(b)$ and $Q_1(b)$ to show the power-law tail with the same exponent $\beta$, based on empirical findings~\cite{Karsai2012Universal}. For this, we adopt the bursty-get-burstier (BGB) merging method~\cite{Jo2017Modeling}, which has been suggested for implementing power-law burst size distributions at various timescales. Then, by considering $Q_0(b)$ as
\begin{equation}
    \label{eq:Q0b}
    Q_0(b) = \zeta(\beta)^{-1} b^{-\beta}\ \textrm{for}\ b=1,2,\cdots,
\end{equation}
one can derive 
\begin{equation}
    \label{eq:Q1b}
    Q_1(b) = \zeta(\beta)^{-1} \left(\frac{b}{s}\right)^{-\beta}\ \textrm{for}\ b=s, 2s,\cdots,
\end{equation}
where $s$ is an integer larger than $1$, corresponding to the average number of bursty trains in $B_0$ per bursty train in $B_1$. For understanding this, we shortly introduce the BGB method: The burst sizes in $B_0$ are sorted in an ascending order. Then the smallest $s$ bursts in $B_0$ are merged into one burst in $B_1$. Then the next smallest $s$ bursts are merged into another burst in $B_1$. In such a way, the $s$ bursts in $B_0$ are sequentially merged into each burst in $B_1$ until all bursts in $B_0$ are used up. In almost all cases, bursts of size $b$ in $B_0$ are merged into bursts of size $sb$ in $B_1$, explaining why all $b$ for $Q_1(b)$ are to be multiples of $s$. It is also found that $Q_1(b)$ shows the same power-law tail as in $Q_0(b)$, as demonstrated for a wide range of the exponent value in Ref.~\cite{Jo2017Modeling}. Accordingly, since a burst of size $s$ in $B_1$ consists of $s$ bursts of size $1$ in $B_0$, we can obtain $t_{11}$ as
\begin{equation}
    \label{eq:t11}
    t_{11} \simeq \frac{(s-2)Q_1(s)}{\langle b_1\rangle},
\end{equation}
and we also find $Q_1(1)=0$ in Eqs.~(\ref{eq:d1}--\ref{eq:d3}). Then all other $t_{\mu\nu}$ can be obtained from Eqs.~(\ref{eq:t00_multi}--\ref{eq:t22_multi}). Hence, we can eventually get the analytical solution of the memory coefficient in Eq.~(\ref{eq:memory_approx_multi}).

We investigate the dependence of $M$ on $Q_0(b)$ and $Q_1(b)$, while keeping the same $P(\tau)$ in Eq.~(\ref{eq:Ptau_powerlaw}). We first consider the pure power-law case of $P(\tau)$ with $\tau_c\to\infty$, where $\alpha>3$ is assumed for the existence of $\sigma$. From Eq.~(\ref{eq:mn_event_burst_multi}), we obtain the relations between parameters as
\begin{eqnarray}
    \label{eq:relation_param_multi1}
    \langle b_0\rangle &=& \frac{\zeta(\beta-1)}{\zeta(\beta)} \simeq \left(\frac{\Delta t_0}{\tau_{\rm min}}\right)^{\alpha-1},\\
    \label{eq:relation_param_multi2}
    \langle b_1\rangle &=& s \langle b_0\rangle \simeq \left(\frac{\Delta t_1}{\tau_{\rm min}}\right)^{\alpha-1}.
\end{eqnarray}
In order to study the dependence of $M$ on $\beta$, we fix the values of $\alpha$ and $\tau_{\rm min}$ for keeping the same $P(\tau)$, implying that $\langle\tau\rangle$ and $\sigma$ in Eq.~(\ref{eq:memory_approx_multi}) remain the same. Then the variation of $\beta$ affects $\Delta t_0$ by means of Eq.~(\ref{eq:relation_param_multi1}). Using the resultant value of $\langle b_0\rangle$ in Eq.~(\ref{eq:relation_param_multi1}), we obtain the relation between $s$ and $\Delta t_1$ by Eq.~(\ref{eq:relation_param_multi2}). For the determination of $\Delta t_1$, we can set a proper value of $s$, e.g., $s=4$. Once $\Delta t_0$ and $\Delta t_1$ are determined, we calculate $\tau^{(\mu)}$ for $\mu=0,1,2$ in Eqs.~(\ref{eq:tau01_mean_multi}) and~(\ref{eq:tau2_mean_multi}) to get the analytical solution of the memory coefficient. This analytical result of $M$ is plotted for various values of parameters, as depicted by solid lines in Fig.~\ref{fig:multi}(a). We find the qualitatively same results as in the single timescale analysis, such as the overall decreasing behavior of $M$ as a function of $\beta$ and the slightly increasing behavior of $M$ for sufficiently large values of $\alpha$. 

Interestingly, $M$ turns out to be negative for large values of $\beta$. We find that $M$ remains negative for very large $\beta$ as shown in Fig.~\ref{fig:multi}(b), which seems to be contradictory with a plausible intuition that the infinite $\beta$ corresponds to the case with uncorrelated interevent times, i.e., $M=0$. For sufficiently large values of $\beta$, one finds $\langle b_0\rangle\approx 1$, implying that almost all bursts in $B_0$ are of size $1$ when using $\Delta t_0$. This may lead to the uncorrelated interevent times as in the single timescale analysis, which is the reason why $M$ approaches $0$ for the increasing $\beta$ in Fig.~\ref{fig:single}. In contrast, in the multiple timescale analysis, the value of $s > 1$ can introduce anti-correlations between interevent times when $\beta$ is very large: In the limiting case with $\beta\to\infty$, if $s=2$, every burst in $B_1$ has the size of $2$ because every burst in $B_0$ has the size of $1$. Accordingly, odd-numbered interevent times are smaller than $\Delta t_1$, while even-numbered interevent times are larger than $\Delta t_1$. This explains the negativity of $M$. Then the larger $s$ is expected to result in the smaller anti-correlations between interevent times, hence the value of $M$ closer to $0$. We confirm this expectation, e.g., for $\alpha=3.5$ as shown in Fig.~\ref{fig:multi}(b). Finally, it is also found that the smaller $\alpha$ leads to the smaller variation of $M$ for fixed $\beta$ and $s$, which can be understood by the fact that the large values of $\Delta t_0$ and $\Delta t_1$ due to the small $\alpha$ enhance the mixing of interevent times with various timescales at least within $T_0$.

Next, we consider the general form of $P(\tau)$ in Eq.~(\ref{eq:Ptau_powerlaw}) with finite $\tau_c$ for the wider range of $\alpha$. The calculation of $M$ is again straightforward and the analytical result of $M$ for various values of parameters is depicted as solid lines in Fig.~\ref{fig:multi}(c). Similarly to the results in the single timescale analysis, we find the overall decreasing behavior of $M$ as a function of $\beta$, as well as the non-monotonic behavior of $M$ according to $\alpha$. We also find the negative $M$ for the wide range of $\beta$.

\subsection{Numerical demonstration}\label{subsec:multi-numeric}

In order to numerically demonstrate the effect of the burst size distribution on the memory coefficient, we construct the sequence of correlated interevent times from the same $T$ as in the previous Section but by means of the BGB method~\cite{Jo2017Modeling} as described in the previous Subsection. Once the sequence of interevent times is obtained, we immediately calculate the value of $M$ in Eq.~(\ref{eq:memory_original}).

The numerical results of $M$ as a function of $\beta$ for various values of $\alpha$ are shown in Fig.~\ref{fig:multi}(a,c), where each point is obtained from $50$ event sequences of size $n=5\times 10^5$. For both cases with and without exponential cutoffs for $P(\tau)$ in Eq.~(\ref{eq:Ptau_powerlaw}), we find that the numerical results are comparable to the analytical values of $M$. However, systematic deviations are observed in the pure power-law case, especially for small values of $\alpha$ and $\beta$, where the natural cutoffs of power-law distributions, i.e., $P(\tau)$ and/or $Q_l(b)$ for $l=0,1$, become more effective due to the finite sizes of $n$ and/or $m_l$ for $l=0,1$. 

\subsection{Solvability of the general case with more than two timescales}\label{subsec:multi-general}

In general, if we consider $k$ time windows with $k\geq 1$, the set of interevent times, $T$, can be divided into $k+1$ subsets. This implies that the number of $t_{\mu\nu}$ for $\mu,\nu\in \{0,1,\cdots, k\}$ is $(k+1)(k+2)/2$ by assuming that $t_{\mu\nu}=t_{\nu\mu}$, while the number of relations for them is $2k+1$, as we have one normalization condition for $t_{\mu\nu}$, i.e., $\sum_{\mu,\nu}t_{\mu\nu}=1$, and two relations for each time window. Therefore, for our single timescale analysis, corresponding to $k=1$, all $t_{\mu\nu}$ could be fully determined only in terms of $Q_0(b)$. In contrast, for the general case with $k\geq 2$, $t_{\mu\nu}$ cannot be fully determined in terms of $Q_l(b)$ for $l=0,1,\cdots,k-1$. However, as we have shown above, one can exploit detailed information on the relations between $Q_l(b)$ to extract more relations between $t_{\mu\nu}$ and/or $Q_l(b)$, e.g., the relation between $t_{11}$ and $Q_1(b)$ in Eq.~(\ref{eq:t11}).

\section{Conclusion}\label{sec:concl}

Temporal inhomogeneities in event sequences of natural and social phenomena have been characterized in terms of interevent times and correlations between interevent times. For the last decade, the statistical properties of interevent times have been extensively studied, while the correlations between interevent times, often called correlated bursts, have been largely unexplored. For measuring the correlated bursts, two relevant approaches have been suggested, i.e., memory coefficient~\cite{Goh2008Burstiness} and burst size distribution~\cite{Karsai2012Universal}. While the memory coefficient $M$ measures correlations between two consecutive interevent times, the burst size distribution can measure correlations between an arbitrary number of interevent times. Recent empirical analyses have shown that burst size distributions follow a power law with exponent $\beta$ for a wide range of timescales~\cite{Karsai2012Universal, Karsai2012Correlated, Wang2015Temporal}, implying the existence of hierarchical burst structure. We observe a tendency that the larger value of $M$ is associated with the smaller value of $\beta$. In addition, empirical findings in human activity patterns appear inconsistent, such that the values of $M$ are close to $0$, while burst size distributions follow a power law with exponent $\beta\approx 4$ for a wide range of time windows for detecting bursty trains. 

As little is known about the relation between memory coefficient and burst size distribution, we have studied their relation by deriving the analytical form of the memory coefficient as a function of parameters describing the interevent time and burst size distributions. For this we have assumed the conditional independence between consecutive interevent times for the sake of analytical treatment. We could demonstrate the general tendency of smaller values of $\beta$ leading to the larger values of $M$, both analytically and numerically. We could also explain why apparently inconsistent observations have been made in human activities: The negligible $M$ turns out to be compatible with power-law burst size distributions with $\beta\approx 4$. Hence, we raise an important question regarding the effectiveness or limits of $M$ in measuring correlated bursts. Although the definition of $M$ is straightforward and intuitive, it cannot properly characterize the complex structure of correlated bursts in some cases. For overcoming the limits of $M$, one can consider the generalized memory coefficients~\cite{Goh2008Burstiness}, defined as
\begin{equation}
    M_k\equiv \frac{ \langle \tau_i\tau_{i+k}\rangle- \langle \tau_i\rangle \langle \tau_{i+k}\rangle }{\sigma_i\sigma_{i+k} },
\end{equation}
where $\langle \tau_i\rangle$ ($\langle\tau_{i+k}\rangle$) and $\sigma_i$ ($\sigma_{i+k}$) denote the average and standard deviation of interevent times except for the last (the first) $k$ interevent times, respectively. The relations between $\{M_k\}_{k=1,2,\cdots}$ and burst size distributions for a wide range of timescales can be studied in the future for better understanding the correlated bursts observed in various complex systems.

Finally, we briefly discuss generative modeling approaches for the correlated bursts. In this paper we have assumed power-law burst size distributions based on the empirical findings, while it is important to understand the generative mechanisms behind such power-law behaviors. Regarding this issue, to our knowledge, we find only a few modeling approaches, such as two-state Markov chain~\cite{Karsai2012Universal} and self-exciting point processes with a power-law kernel~\cite{Jo2015Correlated}, where the power-law kernel can be related to the Omori's law in seismology~\cite{Lippiello2007Dynamical, deArcangelis2016Statistical}. Although these models successfully reproduce some empirical findings, including the power-law burst size distributions, more work needs to be done for the better understanding as the generative mechanisms for power-law burst size distributions are largely unexplored.

\begin{acknowledgments}
    The authors acknowledge financial support by Basic Science Research Program through the National Research Foundation of Korea (NRF) grant funded by the Ministry of Education (2015R1D1A1A01058958).
\end{acknowledgments}

%\appendix

\bibliographystyle{apsrev4-1}

%\bibliography{/Users/h2jo/Research/_papers/h2jo-papers}
\end{document}